\documentclass[10pt, oneside]{article}   	% use "amsart" instead of "article" for AMSLaTeX format
\usepackage{graphicx}				% Use pdf, png, jpg, or eps§ with pdflatex; use eps in DVI mode
\usepackage{amsmath}    % need for subequations
\usepackage{verbatim}   % useful for program listings
\usepackage{color}      % use if color is used in text
\usepackage{subfigure}  % use for side-by-side figures
\usepackage{hyperref}   % use for hypertext links, including those to external documents and URLs
\usepackage{amssymb}
\usepackage{ulem}
\usepackage{authblk}
\begin{document}

\title{High brightness ultrafast Transmission Electron Microscope based on a laser-driven cold-field emission source: principle and applications}

\author[1]{G.M. Caruso}
\author[,1]{F. Houdellier\thanks{Corresponding author: florent.houdellier@cemes.fr}}
\author[,1]{S. Weber}
\author[2]{M. Kociak}
\author[,1]{A. Arbouet\thanks{Corresponding author: arnaud.arbouet@cemes.fr}}

%\email{arnaud.arbouet@cemes.fr}

\affil[1]{CEMES-CNRS, Universit\'{e} de Toulouse, Toulouse, France}
\affil[2]{Laboratoire de Physique des Solides, B\^{a}timent 510, UMR CNRS 8502, Universit\'{e} Paris Sud, Orsay 91400, France}
%\cortext[cor]{arnaud.arbouet@cemes.fr, florent.houdellier@cemes.fr}
%\fntext[fn1]{This is the specimen author footnote.}
%\fntext[fn2]{Another author footnote, but a little more longer.}

%\date{}							% Activate to display a given date or no date

\begin{abstract}
We report on the development of an ultrafast Transmission Electron Microscope based on a laser-driven cold-field emission source.
We first describe the instrument before reporting on numerical simulations of the laser-driven electron emission.
These simulations predict the temporal and spectral properties of the femtosecond electron pulses generated in our ultrafast electron source.
We then discuss the effects that contribute to the spatial, temporal and spectral broadening of these electron pulses during their propagation from the electron source to the sample and finally to the detectors of the electron microscope.
The spectro-temporal properties are then characterized in an electron/photon cross-correlation experiment based on the detection of electron energy gains. 
We finally illustrate the potential of this instrument for ultrafast electron holography and ultrafast electron diffraction.
\end{abstract}

%\pacs{XXXX.XX}
%\keywords{EELS, Green Dyadic Method, Cathodoluminescence, surface plasmon}

\maketitle
%\tableofcontents 

Thanks to their charge, electrons can be easily accelerated and focused by electric and magnetic fields and interact strongly with matter. 
This makes them an ideal probe of the structure of materials used in a variety of scientific tools such as electron microscopes, synchrotrons or free electron lasers for instance.
Among these instruments, Transmission Electron Microscopes have a unique position due to their ability to provide information in real and reciprocal space as well as in the energy domain \cite{williams_transmission_1996}.
This position has been further reinforced by recent technological advances such as the development of aberration correctors and monochromators that pushed their spatial and spectral resolution to the sub-angstr\"{o}m \cite{haider_electron_1998,krivanek_atom-by-atom_2010} and meV range \cite{krivanek_vibrational_2014} respectively.
Nevertheless, obtaining information in the time domain has long remained a challenge due to the difficulty to generate electron pulses with sub-picosecond durations.
Pioneered in the late 1980s by O. Bostanjoglo and coworkers at the Technical University of Berlin, the field of Time-resolved Transmission Electron Microscopy has been boosted by the spectacular improvement in spatio-temporal resolution achieved in 2005 in the group of A. Zewail at the California Institute of Technology \cite{domer_high-speed_2003,king_high-speed_2007,Zewailbook,zewail_four-dimensional_2010,flannigan_4d_2012,vanacore_four-dimensional_2016,arbouet_chapter_2018}.
Contrary to Dynamic Transmission Electron Microscopes (DTEM) which use single pulses containing each a very large number of electrons, Ultrafast Transmission Electron Microscopes (UTEM) rely on stroboscopic observations with electron pulses each containing only a few particles \cite{lobastov_four-dimensional_2005, piazza_design_2013, cao_clocking_2015, bucker_electron_2016}.
This operation in the single electron regime cancels the coulombic interparticle repulsion that deteriorates the resolution of DTEMs \cite{Gahlmann2008}.
The generation of femtosecond electron pulses and their synchronization with ultrashort light pulses has allowed for instance to manipulate electron wavepackets by exploiting coherent electron/photon interactions  \cite{Feist2015,echternkamp_ramsey-type_2016, priebe_attosecond_2017,vanacore_attosecond_2018,morimoto_diffraction_2018}.
The development of UTEMs enabled many spectacular applications in nanomechanics, nano-optics and nano-magnetism but the low brightness of their electron sources precluded their use for applications such as electron holography for instance \cite{Zewailbook}.
A little more than a decade ago, laser-driven emission of electron pulses from sharp metallic tips has been demonstrated \cite{hommelhoff_ultrafast_2006, Hommelhoff2006a, ropers_ultrafast_2007, ropers_localized_2007}.
Important efforts have since been devoted to the theoretical and experimental investigation of the temporal and spectral properties of the ultrashort electron pulses delivered by tip-based electron sources \cite{Paarmann2012, Ehberger2015,cook_coulomb_2016, bach_coulomb_2019}.
Nanosized electron sources have then been used in electron guns \cite{hoffrogge_tip-based_2014, Bormann2015} and scanning electron microscopes \cite{yang_scanning_2010}.
A UTEM based on a Schottky electron source has been developed in the University of G\"{o}ttingen \cite{Feist2015,Feist2017}.
We have recently developed a UTEM based on a laser-driven cold-field emission source and demonstrated its potential for electron holography \cite{Caruso2017,houdellier_development_2018}.

In the following, we first describe this instrument before reporting on numerical simulations of the laser-driven electron emission process.
Finally we discuss the new possibilities in time-resolved diffraction and holography opened by the improved brightness and spatial coherence of the most recent ultrafast Transmission Electron Microscopes.

%%%%%%%%%%%%%%%%%%%%%%%%%%%%%%%%%%%%%%%%%%%%%%%%%
%%%%%%%%%%%%%%%%%%%%%%%%%%%%%%%%%%%%%%%%%%%%%%%%%
\section{Development of an ultrafast TEM based on a laser-driven cold-field emission source}
%%%%%%%%%%%%%%%%%%%%%%%%%%%%%%%%%%%%%%%%%%%%%%%%%
%%%%%%%%%%%%%%%%%%%%%%%%%%%%%%%%%%%%%%%%%%%%%%%%%

The ultrafast TEM developed in our laboratory is based on a laser-driven cold-field emission source.
The architecture of this electron source and its potential for ultrafast electron microscopy and holography has been detailed in earlier publications \cite{arbouet_chapter_2018,Caruso2017,houdellier_development_2018}.
%%%%%%%%%%%%%%%%%%%%%%%%%%%%%%%%%%%%%%%%%%%%%%%%%
%%%%%%%%%%%%%%%%%%%%%%%%%%%%%%%%%%%%%%%%%%%%%%%%%
\begin{center}
\begin{figure}[htbp]
\centering
\includegraphics[width=12cm,angle =0.]{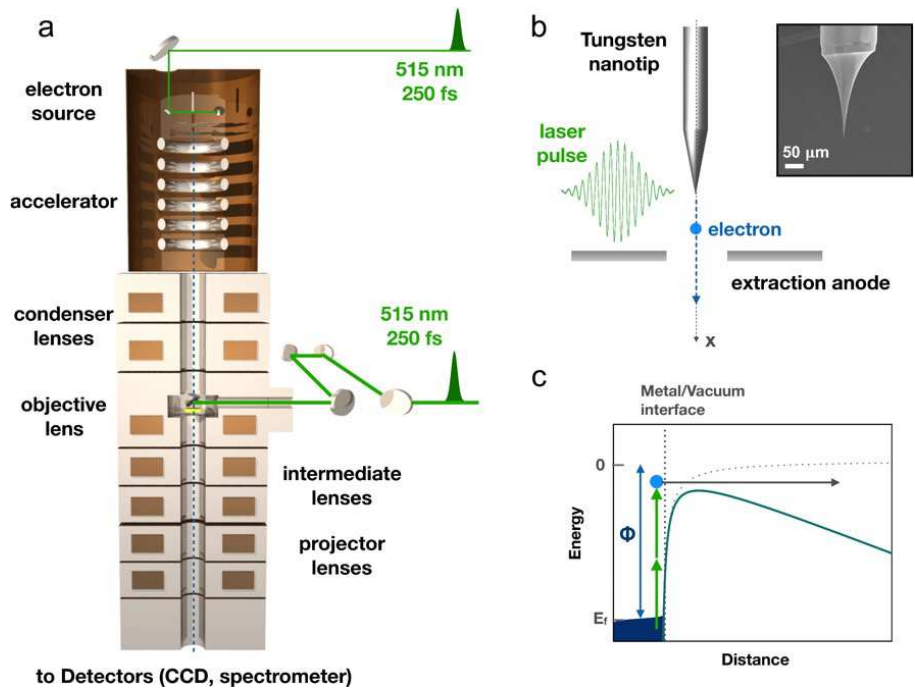}
\caption{(Color Online) a) Ultrafast Transmission Electron Microscope based on a laser-driven cold-field emission source.
b) Close-up of the tip region: the [310] oriented tungsten nanotip is placed in front of an extraction anode and illuminated by femtosecond laser pulses (250 fs, 515 nm).
c) Potential of an electron at the metal/vacuum interface. The potential is lowered by the \textit{Schottky} effect due to the applied extraction voltage $V_{DC}$.
In our experimental conditions, electron emission is triggered by the absorption of two photons ($E_{phot} =2.41$ eV).}
\label{Figure1}
\end{figure}
\end{center}
%%%%%%%%%%%%%%%%%%%%%%%%%%%%%%%%%%%%%%%%%%%%%%%%%
%%%%%%%%%%%%%%%%%%%%%%%%%%%%%%%%%%%%%%%%%%%%%%%%%
As sketched in Figure \ref{Figure1}-a, our UTEM combines an optical pump-probe set-up with a TEM.
The latter is a Hitachi High Technologies 200 keV FE-TEM HF2000 modified to allow light injection inside the objective lens to excite the sample and inside the electron source to trigger the emission of femtosecond electron pulses.
Our optical set-up is based on an amplified fiber laser yielding up to 20 $\mu$J, sub-300 fs pulses at 1030 nm with a repetition rate adjustable between single shot and 40 MHz.
The output of the laser is frequency-doubled in a BBO non-linear crystal before being sent to the electron source.
The power, polarization and position of the laser spot on the tungsten nanotip can be remotely controlled by opto-mechanical components integrated in a so-called \textit{optical head} installed as an add-on above the electron source of the TEM (not shown on the figure for simplicity) \cite{Caruso2017}.
The laser beam is focused by a parabolic mirror (f = 8 mm) installed inside the electron gun in ultra-high vacuum on a commercial [310] oriented monocrystalline tungsten nanotip similar to the ones used in the original HF2000.
In Figure 1-b), the nanotip is placed in front of an extraction anode.
An extraction voltage $V_{DC}=4$ kV is applied between the nanotip and the extraction anode in the experiments reported below. 
This value of $V_{DC}$ is too small to trigger the emission in the absence of optical excitation by the visible laser pulses (250 fs, 515 nm).
The electrons are then accelerated in the accelerator located below the extraction anode.
In the experiments reported below, the acceleration voltage is 150 kV.
To allow excitation of the sample by a laser beam or collection of its cathodoluminescence, a light injection/cathodoluminescence system has been installed in the objective lens.
It is composed of a parabolic mirror placed above the sample which can be adjusted from outside the TEM column by micrometer screws \cite{das_withdrawn:_2018}.

We describe in the following the electron emission process and compute the temporal and spectral properties of the electron pulses.

%%%%%%%%%%%%%%%%%%%%%%%%%%%%%%%%%%%%%%%%%%%%%%%%%
%%%%%%%%%%%%%%%%%%%%%%%%%%%%%%%%%%%%%%%%%%%%%%%%%
\section{Electron emission mechanism, temporal and spectral properties of the femtosecond electron pulses}
%%%%%%%%%%%%%%%%%%%%%%%%%%%%%%%%%%%%%%%%%%%%%%%%%
%%%%%%%%%%%%%%%%%%%%%%%%%%%%%%%%%%%%%%%%%%%%%%%%%

\subsection{Description of the laser-driven ultrafast electron emission}

Our numerical simulations follow a similar approach as the one reported in \cite{wu_nonequilibrium_2008,yanagisawa_laser-induced_2010,yanagisawa_energy_2011,yanagisawa_delayed_2016}.
We consider a gas of free electrons interacting with a femtosecond laser pulse of duration $\Delta t_{laser}$ and a central wavelength $\lambda_p$.
The interface between the metallic system ($x<0$) and the vacuum ($x>0$) is assumed to be flat.
The metallic nanotip is placed in an auxiliary DC electric field $E_{DC}$.
%In our experiments, the latter is kept at a sufficiently low level so that no DC current is detected when the laser beam is not incident on the %nanotip.
%%%%%%%%%%%%%%%%%%%%%%%%%%%%%%%%%%%%%%%%%%%%%%%%%
%%%%%%%%%%%%%%%%%%%%%%%%%%%%%%%%%%%%%%%%%%%%%%%%%
\begin{center}
\begin{figure}[htp]
\centering
\includegraphics[width=12cm,angle =0.]{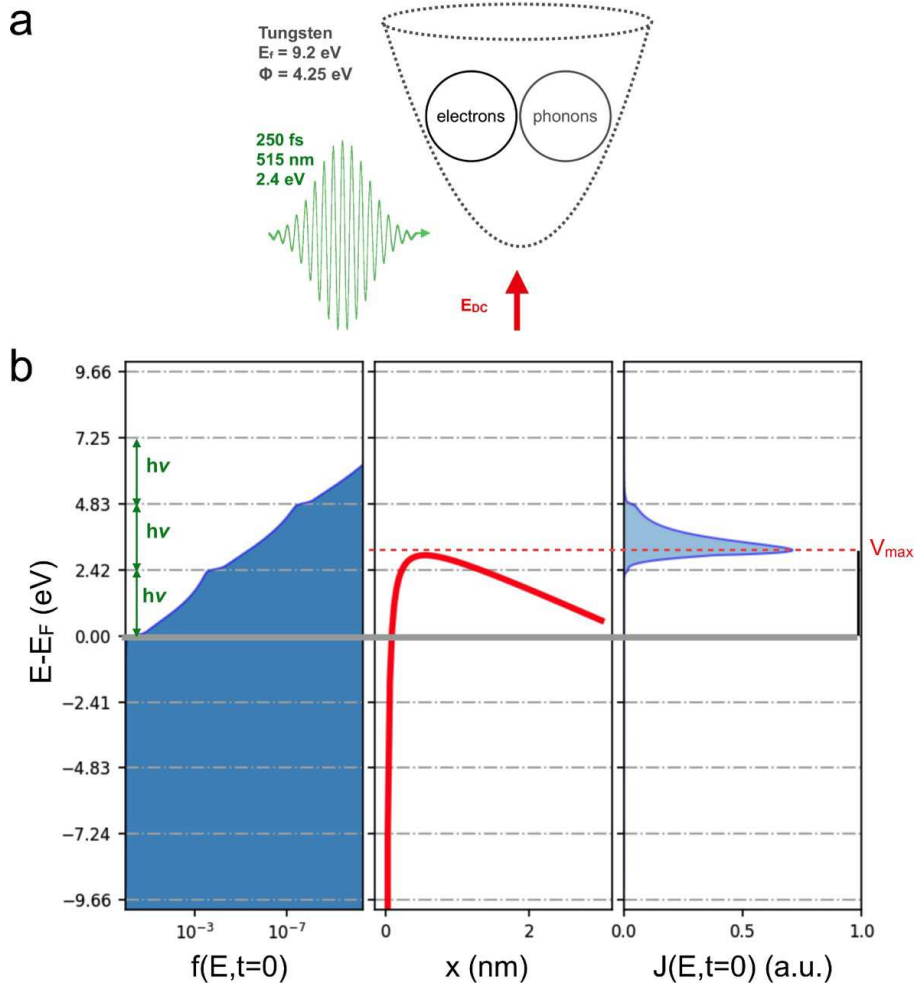}
\caption{(Color Online) a)  A femtosecond laser pulse is incident on a metallic nanotip: the ultrafast dynamics of the excited electron gas involves optical excitation by the laser pulse, electron-electron and electron-phonon interactions on a femtosecond timescale. The laser pulse intensity is maximum on the nanotip at t=0 fs.
b) Left: Electron occupation number as a function of energy at t = 0 fs. 
Center: electron potential at the metal/vacuum interface. 
Right: Instantaneous spectral density of emitted current $J(E, t = 0 \; \mathrm{fs})$.
$V_{max} = 2.94$ eV is the maximum of the potential barrier.}
\label{Figure2}
\end{figure}
\end{center}
%%%%%%%%%%%%%%%%%%%%%%%%%%%%%%%%%%%%%%%%%%%%%%%%%
%%%%%%%%%%%%%%%%%%%%%%%%%%%%%%%%%%%%%%%%%%%%%%%%%
%Figure 1-b shows the potential $V(x)$ of an electron close to a metal/vacuum interface located at $x=0$.
The potential $V(x)$ of an electron can be written:
\begin{equation}
V(x) = \Phi  - \frac{e^2}{16 \pi \varepsilon_0 x^2} - e \, E_{DC} \, x 
\end{equation}
in which $\Phi$ is the workfunction of the metal and $\varepsilon_0$ is the vacuum permittivity.
The second term is the contribution from the image charge induced inside the metal by the emitted electron.
The last term accounts for the additional influence of the external DC electric field $E_{DC}$ which lowers the potential barrier by the \textit{Schottky} effect.
$E_{DC}$ has been computed using the Finite-Difference Method-based software SIMION and cross-checked using the  finite element method (FEM)
software COMSOL multiphysic \cite{Caruso2017}.
The computed electric field at the tip apex for an extraction voltage $V_{DC} = 4$ kV is 1.19 V/nm.
This value is lower than in the original electron source of the HF2000 due to the presence of the mirror holder which partially screens the electric field.
This explains why no electron emission is obtained in the absence of optical excitation.
As sketched in Figure 1-c, the applied $E_{DC}$ lowers the potential barrier by $\Delta \Phi_{schottky} = 1.3$ eV so that electron emission can only be triggered by the absorption of at least 2 photons with energy $E_{phot}= 2.41$ eV.

The number of electrons emitted at time $t$ per unit area and unit time with a total energy between $E$ and $E+dE$ from the metal surface can be written as:
\begin{equation}
N(E,t) = \int_0^E N(E,W,t) D(W,E_{DC}) dW
\end{equation}
 $N(E,W,t)$ is the number of electrons with a total energy between $E$ and $E+dE$ and normal energy with respect to the tip surface between $W$ and $W+dW$.
The total energy $E$ can be written as a function of the electron wavevector components $(k_x , k_y ,k_z)$ as $E = \hbar^2 (k_x^2 +k_y^2+k_z^2)/2m$ whereas the contribution to the kinetic energy normal to the interface is $W = \hbar^2 k_x^2 /2m$. 
$D(W,E_{DC})$ is the probability that an electron with a normal energy $W$ tunnels through the potential barrier.
The transmission coefficient $D(W,E_{DC})$ is computed in the \textit{Wentzel-Kramers-Brillouin} approximation \cite{wu_nonequilibrium_2008,yanagisawa_laser-induced_2010,yanagisawa_energy_2011}.
$N(E,W,t)$ can be expressed in the framework of the free-electron theory of metals as:
\begin{equation}
N(E,W,t) = \frac{m}{2 \pi^2 \hbar^3} f(E,t) dW dE
\end{equation}

The spectral density of emitted electrons at time $t$ can be written as:
\begin{equation}
J(E,t) = - e \int_0^{E} N(E,W,t) dW
\label{JET}
\end{equation}

The total current emitted at time $t$ can be written as :
\begin{equation}
J(t) = - e \int_0^{+\infty} \int_0^{E} N(E,W,t) \, dW \, dE
\label{JT}
\end{equation}

The spectrum of the electron pulse is obtained after integration on the pulse duration as:
\begin{equation}
J(E) = e \int_{-\infty}^{+\infty} \int_0^{E} N(E,W,t) \, dW \, dt
\label{JE}
\end{equation}

The computation of the spectral and temporal distribution of the emitted electrons is done in two steps.
First, the time-dependent electron distribution $f(E,t)$ is calculated.
The electron pulse spectrum $J(E)$, temporal profile $J(t)$ and spectral density $J(E,t)$ at time $t$ are then obtained from the distribution of the electrons on the different energy levels using equations \ref{JET}, \ref{JT} and \ref{JE}.

As sketched in Figure \ref{Figure2}-a), the description of the ultrafast dynamics of the tungsten nanotip upon excitation by a femtosecond laser pulse must take into account the optical excitation by the laser pulse as well as the electron-electron and electron-phonon scattering processes.
The temporal evolution of the electron occupation number $f(\textbf{k},t)$ is then computed from the Boltzmann equation:
\begin{equation}
\frac{df(E,t)}{dt} = \left.\frac{df(E,t)}{dt}\right|_{e-e} + \left.\frac{df(E,t)}{dt}\right|_{e-ph} +  O(E,t)
\end{equation}
In the latter, $O(E,t)$ accounts for the optical excitation of the electron gas by the laser pulse.
We assume an isotropic conduction band with a dispersion $E({\bf k}) = E(k) = \hbar^2 k^2/ 2m$ and a Fermi energy $E_F =9.2$ eV (the origin of energies is taken at the bottom of the conduction band).
Therefore, only a dependence of the occupation number upon time and electron energy is kept in the following.
The workfunction of tungsten is taken as $\Phi = 4.25$ eV.
Initially, electrons and phonons are thermalized and described by Fermi-Dirac and Bose-Einstein distributions respectively.
The optical excitation of the electron gas as well as electron-electron and electron-phonon interactions are taken into account at each time step of the computation.
The corresponding  collision terms are taken from \cite{del_fatti_nonequilibrium_nodate,wu_nonequilibrium_2008,yanagisawa_energy_2011}.
The strength of electron-phonon interactions is adjusted to yield an electron-phonon thermalization time in agreement with the results of time-resolved pump-probe spectroscopy experiments \cite{fujimoto_femtosecond_1984}.
The Botzmann equation is numerically resolved with typical energy steps of 2 meV and time steps of 1 fs.

The magnitude of the optical excitation is quantified by the excess energy per unit volume $\Delta u_{exc}$ injected inside the electron gas by the laser pulse.
The latter has been estimated using electrodynamical simulations \cite{wiecha_pygdmpython_2018}.
The apex of the nanotips used in our electron source has a peculiar shape which can be approximated by a nanosphere (see Figure 2-d in \cite{houdellier_development_2018}).
In our experiments, the average laser power sent inside the electron gun is in the range 4-8 mW at a laser repetition rate 2 MHz.
Our laser beam is focused in a spot having a diameter of $\sim$ 6 $\mu m$ ($1/e^2$  intensity) yielding a peak power on the tip in the $10^{10}-10^{11}$ $\mathrm{W/cm}^2$ range.
Our calculations show that the average power absorbed in the tip apex is of the order of a few $\mu W$ and that the energy density injected in the electron gas is close to 1 $\mathrm{eV/nm^3}$ per laser pulse.
An estimation of the steady-state temperature increase shows that it is only of a few Kelvins: this rules out any contribution from laser-induced thermionic emission.
 
%%%%%%%%%%%%%%%%%%%%%%%%%%%%%%%%%%%%%%%%%%%%%%%%%
%%%%%%%%%%%%%%%%%%%%%%%%%%%%%%%%%%%%%%%%%%%%%%%%%
%\begin{center}
%\begin{figure}[htp]
%\centering
%\includegraphics[width=13cm,angle =0.]{Figure3.eps}
%\caption{(Color Online) Left: electronic occupation number at $t=0$ fs, time at which the incident laser pulse intensity is maximum on the metal. Center: Potential $V(x)$ as a %function of distance to metal/vacuum interface $x$. Right: Spectral density of emitted current $j(E,t)$ at $t=0$ fs.}
%\label{Figure3}
%\end{figure}
%\end{center}
%%%%%%%%%%%%%%%%%%%%%%%%%%%%%%%%%%%%%%%%%%%%%%%%%
%%%%%%%%%%%%%%%%%%%%%%%%%%%%%%%%%%%%%%%%%%%%%%%%%
Figure \ref{Figure2}-b shows an example of the simulation results when the incident laser pulse is at its maximum intensity on the metallic system ($t=0$ fs).
These simulations have been performed exactly in our experimental conditions with a photon energy of 2.4 eV, a pulse temporal full width at half-maximum (FWHM) in intensity of 250 fs.
We assume that the optical excitation deposits an excess energy density of 0.4 $\mathrm{eV/nm}^3$ in the electron gas.
The electronic occupation number $f(E,t)$ is represented on a logarithmic scale on the left.
Clear discontinuities in the electronic distribution are visible separated from the Fermi level by multiples of the photon energy.
These discontinuities are associated to the absorption of one, two or three photons by the electrons.
They are smoothed by fast electron-electron interactions which affect the electron distribution even before the laser pulse leaves the metallic medium.
In the center, the electron potential at the interface is represented.
We show on the right the instantaneous emitted current $j(E, t=0 fs)$.
Two contributions are clearly visible.
The first arises from electrons which have gained enough energy from the laser pulse to be promoted to energy levels lying above the top of the potential barrier.
These electrons are directly emitted in vacuum by field-assisted photoemission.
The second contribution comes from electrons which have absorbed two photons and were promoted to energy levels lying slightly below the top of the potential barrier.
Due to the small barrier thickness, these electrons have a high probability of tunneling through the barrier.
This second mechanism is called photo-assisted field-emission.
For an excess energy density of 0.4 $\mathrm{eV/nm}^3$, 88 $\%$ of the electrons are emitted by field-assisted photoemission and 12 $\%$ are emitted by photo-assisted field-emission.
It is worth noting that the fact that these two mechanisms contribute to the emission and their relative contribution is a result of the experimental conditions of our first experiments.
It is possible to modify the relative contribution of these two mechanisms by changing the height of the potential barrier by altering the static $E_{DC}$ electric field or changing the wavelength of the laser pulses.
In the conditions of Figure \ref{Figure2}-b, the number of electrons emitted per laser pulse per unit surface is $2.6\mathrm{x}10^{-4} \mathrm{nm}^{-2}$.
Assuming that electron emission originates from the half-sphere at the end of the emission tip (R$\sim$100 nm), we can estimate that the total number of electrons emitted per laser pulse is $\sim$ 10.
%%%%%%%%%%%%%%%%%%%%%%%%%%%%%%%%%%%%%%%%%%%%%%%%%
%%%%%%%%%%%%%%%%%%%%%%%%%%%%%%%%%%%%%%%%%%%%%%%%%
\begin{center}
\begin{figure}[htp]
\centering
\includegraphics[width=13cm,angle =0.]{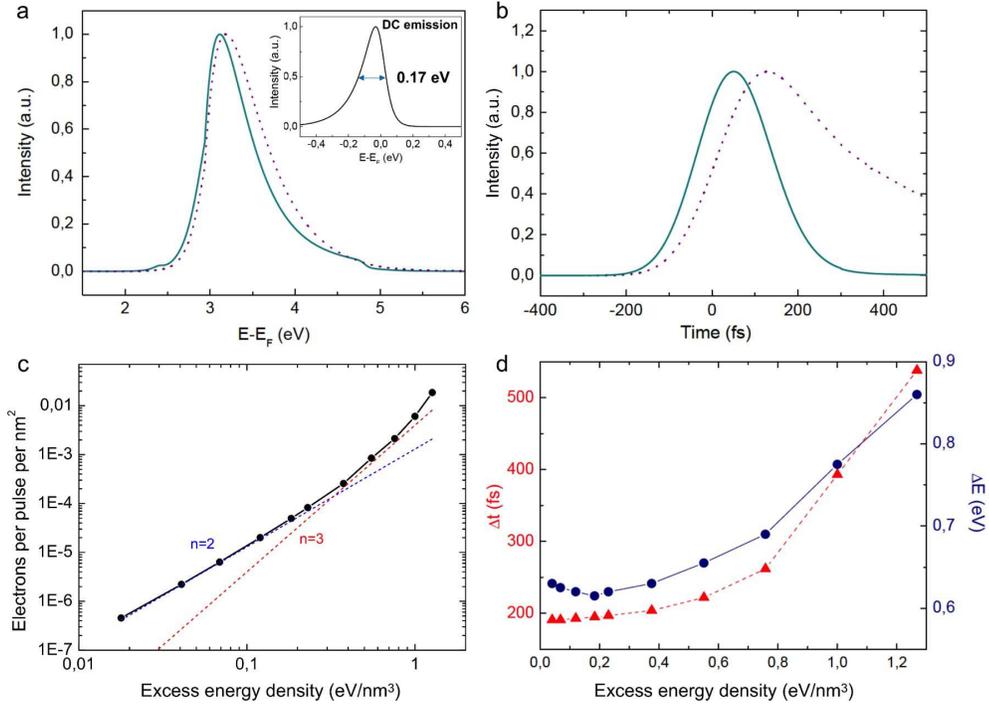}
\caption{(Color Online) a) Spectrum of the electron pulse computed in laser-driven mode. Inset: electron energy spectrum computed for a conventional DC cold-field emission source. 
b) Temporal profile of the emitted electron pulse. The results in a) and b) have been computed for an excess energy density $\Delta u_e = 0.4 \;  \mathrm{eV/nm}^3$ (solid line) and $\Delta u_e = 1 \;  \mathrm{eV/nm}^3$ (dashed line).
c) Number of electrons per pulse per unit surface as a function of the excess energy density.
d) Dependence of the temporal (red triangles) and energy (blue dots) width (FWHM) on the excess energy density.}
\label{Figure4}
\end{figure}
\end{center}
%%%%%%%%%%%%%%%%%%%%%%%%%%%%%%%%%%%%%%%%%%%%%%%%%
%%%%%%%%%%%%%%%%%%%%%%%%%%%%%%%%%%%%%%%%%%%%%%%%%

In Figure 3-a, we show the spectrum computed in laser-driven emission mode ($E_{DC} = 1.19$ V/nm, $E_{phot} = 2.41$ eV) for an excess energy density of 0.4 $\mathrm{eV/nm}^3$.
The energy distribution in these conditions has a FWHM of 0.65 eV.
For reference, we show in inset the emission spectrum computed for conventional (i.e non laser-driven) field-emission from the tungsten nanotip.
The energy distribution has the asymmetric shape characteristic of cold-field emission sources with a FWHM of 0.17 eV.
By comparing the results for normal and laser-driven emission, it is clear that the asymmetry of the spectrum is flipped in one case with respect to the other.
Figure 3-b shows the temporal profile of the laser-driven electron emission which has a FWHM of 204 fs.
To illustrate the influence of the energy density injected in the electron gas, we have computed the emitted current and the temporal and spectral width of the electron pulses for different values of the injected energy density.
Figure 3-c shows that for small perturbations of the electronic system, the emitted current scales as the square of the deposited optical energy density.
As mentionned earlier, this is expected as two photons are required to overcome the potential barrier.
When the number of electrons per laser pulse per unit area increases beyond $\sim \; 10^{-4} \:\mathrm{e/pulse/nm^2}$, i.e for an excess energy density larger than $\sim \; 0.5 \;  \mathrm{eV/nm}^3$, the effective nonlinearity order increases significantly and varies with the injected energy density.
The temporal and spectral properties of the electron pulses degrade in this strong perturbation regime and depend on the optical excitation.
%For instance, $\Delta t$ = 500 fs and  $\Delta E$ = 0.9 eV for $\Delta u_e = 1.2 \; \mathrm{eV/nm}^3$.
In this regime, the influence of the stronger perturbation of the electron distribution is clearly visible both in the time and energy domain.
The spectrum is broadened and the temporal profile is also broadened and becomes asymmetric.
This asymmetry comes from the fact that the rising edge of the electron pulse is related to energy injection inside the electron gas whereas the falling edge at positive times is influenced by the energy transfer from the electron gas to the lattice.
Whereas the former has a characteristic time given by the laser pulse FWHM, the latter increases for strong perturbations of the electron gas \cite{fatti_femtosecond_2006}.
This is clearly visible in Figure 3-a (resp. 3-b) where the dashed lines show the spectrum (resp. temporal profile) computed for an excess energy density of 1 $\mathrm{eV/nm}^3$.
We have performed systematic measurements of the emitted current as a function of the optical excitation and obtained consistently a nonlinearity in the $\sim 3-3.5$ range \cite{houdellier_development_2018}.
According to our simulations, this value would correspond to a pulse duration in the 262-393 fs range and an energy spread between 0.7 and 0.78 eV.
To go further, more precise calculations taking into account the inhomogeneity of the optical near-field at the tip apex and the associated spatial dependence of the electron yield would be required.

%%%%%%%%%%%%%%%%%%%%%%%%%%%%%%%%%%%%%%%%%%%%%%%%%
%%%%%%%%%%%%%%%%%%%%%%%%%%%%%%%%%%%%%%%%%%%%%%%%%
\subsection{Influence of propagation effects on the spectral, spatial and temporal properties of the femtosecond electron pulses}
%%%%%%%%%%%%%%%%%%%%%%%%%%%%%%%%%%%%%%%%%%%%%%%%%
%%%%%%%%%%%%%%%%%%%%%%%%%%%%%%%%%%%%%%%%%%%%%%%%%

To account for the  spectral shapes and temporal profiles measured experimentally in our ultrafast TEM, it is required to take into account the modifications of these spectral and temporal properties during the propagation from the electron source to the interaction region and later to the TEM detector.
Two effects must be considered: the influence of the initial energy spread and space-charge effects.

When the average number of electrons per pulse is significantly smaller than one, the electron-electron repulsive interactions are absent and do not degrade the characteristics of the generated pulses \cite{Zewailbook}.
However, even in this situation the spectro-temporal properties of the electron pulses will be modified during their propagation due to the initial energy spread of the particles.
Indeed, the measured pulse duration results from the statistical averaging of subsequent measurements involving electrons having different initial velocities at the exit of the cathode.
%This dispersion-induced broadening can be estimated by solving the relativistic equation of motion for an electron.
It can be shown that the broadening in the field-free region is negligible \cite{Gahlmann2008}.
The longitudinal broadening $\Delta t_{dis}$ induced by an initial energy spread $\Delta E_i$ is mainly accumulated during the acceleration.
In a region of constant electric field $E_{acc}$, $\Delta t_{dis}$ can be estimated as \cite{Gahlmann2008}:
\begin{equation}
\Delta t_{dis} = \frac{1}{e E_{acc}} \sqrt{\frac{m}{2}}\frac{\Delta E_i}{\sqrt{E_i}}
\label{DTdis}
\end{equation}
However, the direct application of formula \ref{DTdis} is not possible in our case as the electric field has very large gradients around the nanotip \cite{quinonez_femtosecond_2013}.
We have instead computed the transit time from the nanotip to the sample by numerical integration of the relativistic equation of motion of the electrons.
%The acceleration voltage is 150kV as in the experiments described below.
Considering only electrons propagating on-axis from the tip to the sample, we calculated that an initial energy spread $\Delta E_i = 1$ eV translates into a temporal broadening $\Delta t_{dis} = 	120$ fs.
Off-axis trajectories yield larger but comparable values of the transit time.
Considering the different contributions to the electron pulse broadening, the pulse duration can be written as:
\begin{equation}
\Delta t = \sqrt{\Delta t_{em}^2 + \Delta t_{dis}^2} 
\end{equation}
in which $\Delta t_{em}$ is the temporal FWHM associated to the emission process computed in the previous section.
Our simulations show that in the absence of space-charge effects and assuming an electron pulse duration at the cathode $\Delta t_{em} \sim 300$ fs), the minimum electron pulse duration attainable in our UTEM should be close to 350 fs.

%%%%%%%%%%%%%%%%%%%%%%%%%%%%%%%%%%%%%%%%%%%%%%%%%
%%%%%%%%%%%%%%%%%%%%%%%%%%%%%%%%%%%%%%%%%%%%%%%%%
\begin{center}
\begin{figure}[htp]
\centering
\includegraphics[width=9cm,angle =0.]{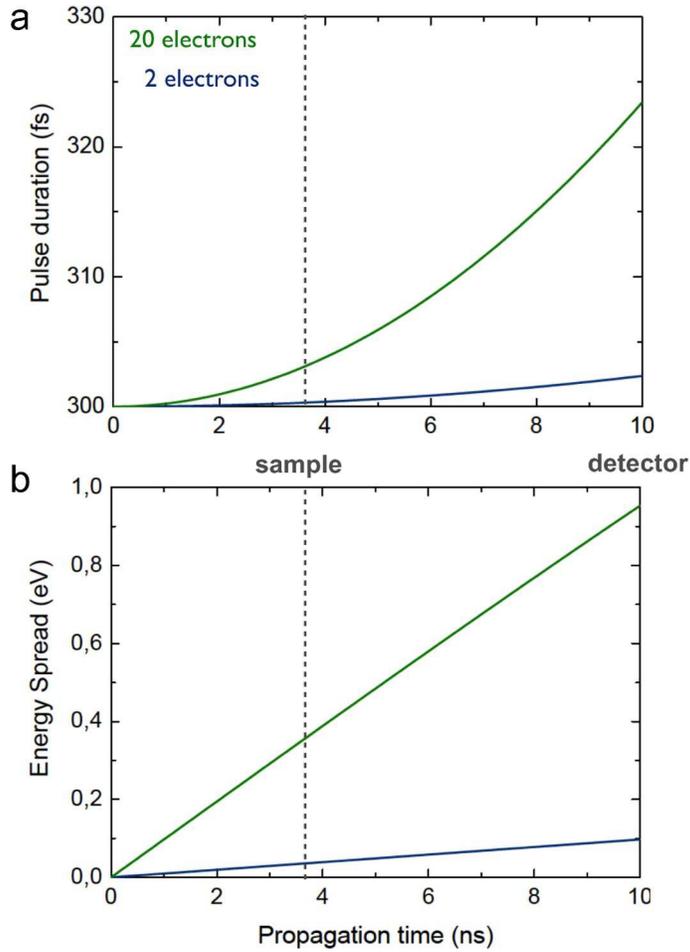}
\caption{(Color Online) Electron pulse duration (a) and energy spread (b) as a function of propagation time for two different numbers of electron per pulse. We assume that the electron pulse has an initial radius of 320 $\mu m$, a kinetic energy of 150 kV and an initial duration of 300 fs.}
\label{FigureSP}
\end{figure}
\end{center}
%%%%%%%%%%%%%%%%%%%%%%%%%%%%%%%%%%%%%%%%%%%%%%%%%
%%%%%%%%%%%%%%%%%%%%%%%%%%%%%%%%%%%%%%%%%%%%%%%%%

In our experiments, the number of electrons per pulse at the tip apex lies between 1 and 20.
Even at these low currents, it is instructive to estimate the influence of space-charge effects during the electron pulse propagation from the source to the sample and detector.
In the following, we adopt a mean-field approach \cite{Gahlmann2008}.
Approximating a pulse of $N_{el}$ electrons as a uniform distribution of charge and considering the force acting on an electron at the leading edge of pulse due to the charge distribution inside the pulse, the temporal and spectral broadening of the electron pulse can be computed for different values of $N_{el}$.
The evolution of the temporal and energy spread along the electron trajectory is shown in Figure \ref{FigureSP}-a-b for an electron pulse having an initial duration of 300 fs electron pulse and a lateral size of 320 $\mu m$.
This estimation of the extent of the electron beam in the transverse direction has been extracted from simulations of the electron trajectories in the TEM column performed with SIMION.
It is an average which does not take into account the presence of cross-overs.
The results displayed in Figure \ref{FigureSP}-a-b are therefore only indicative of the influence of space-charge effects in our microscope.
Nevertheless, they suggest that electron pulses with only a few particles should keep durations in the sub-picosecond range along the trajectory from the tip to the sample.
The energy spread increases from 0.1 to 1 eV when increasing the number of electrons from 2 to 20.
It is important to realize that the validity of such a mean-field approach is questionable for such low number of electrons per pulse.
Recently, the spectral and temporal properties of the electron pulses generated by a laser-driven Schottky electron gun have been computed using a Verlet algorithm and taking into account the intrapulse coulomb interactions, the external electric field and influence of image charges at the emitter.

In the next section we characterize the spectral, spatial and temporal properties of the ultrashort electron pulses in our UTEM by Electron Energy Loss Spectroscopy (EELS) and Electron Energy Gain Spectroscopy (EEGS).

%%%%%%%%%%%%%%%%%%%%%%%%%%%%%%%%%%%%%%%%%%%%%%%%%
%%%%%%%%%%%%%%%%%%%%%%%%%%%%%%%%%%%%%%%%%%%%%%%%%
\section{Characterization of the spatial, temporal and spectral properties of the ultrashort electron pulse}
%%%%%%%%%%%%%%%%%%%%%%%%%%%%%%%%%%%%%%%%%%%%%%%%%
%%%%%%%%%%%%%%%%%%%%%%%%%%%%%%%%%%%%%%%%%%%%%%%%%
%%%%%%%%%%%%%%%%%%%%%%%%%%%%%%%%%%%%%%%%%%%%%%%%%
\subsection{Influence of the number of electrons per pulse on the energy spectrum and spot radius}
%%%%%%%%%%%%%%%%%%%%%%%%%%%%%%%%%%%%%%%%%%%%%%%%%
%%%%%%%%%%%%%%%%%%%%%%%%%%%%%%%%%%%%%%%%%%%%%%%%%
%%%%%%%%%%%%%%%%%%%%%%%%%%%%%%%%%%%%%%%%%%%%%%%%%
\begin{center}
\begin{figure}[htbp!]
\centering
\includegraphics[width=9cm,angle =0.]{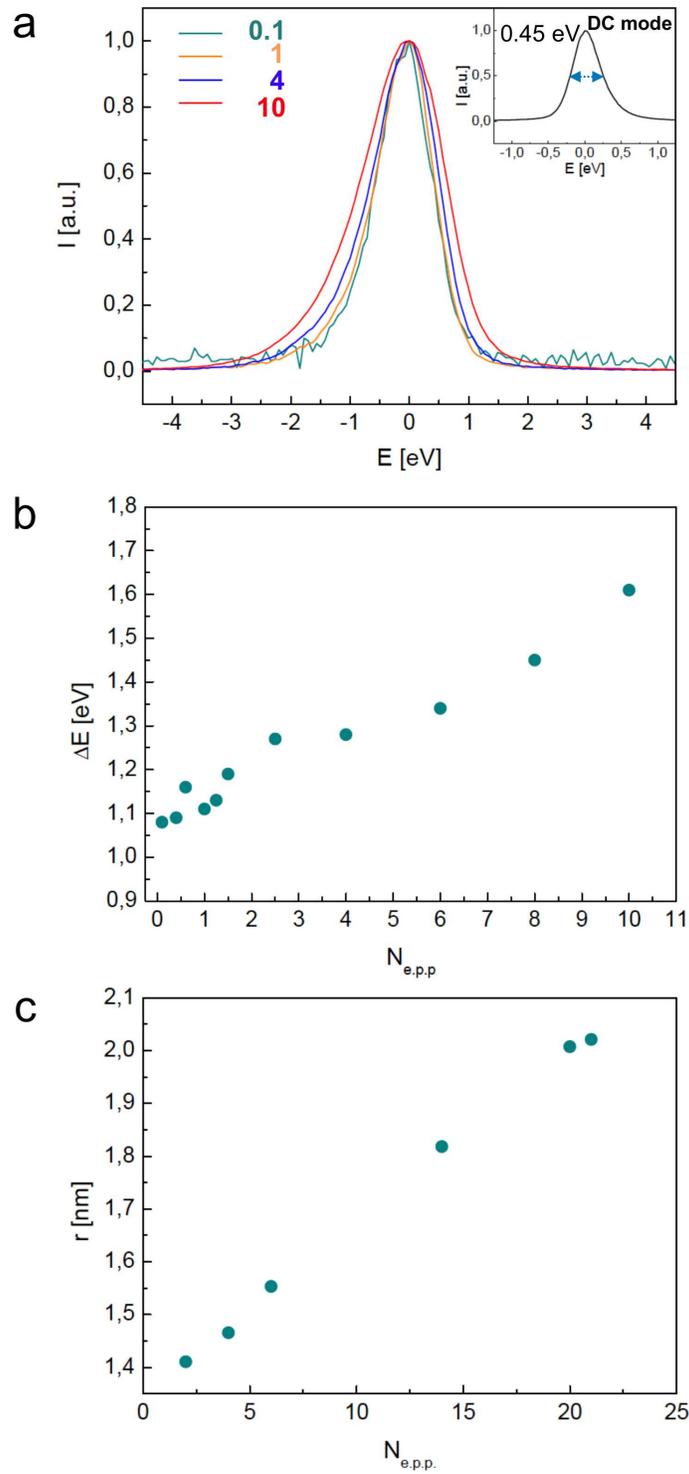}
\caption{(Color Online) a) Zero-Loss Spectrum of the electron pulse in laser-driven mode for different number of electrons per pulse.
Inset: Zero-loss spectrum measured in DC mode.
b) FWHM of the electron energy spectrum as a function of the number of electrons per pulse.
c) Minimum achievable electron spot size (FWHM) as a function of the number of electrons per pulse.}
\label{FigureZLP}
\end{figure}
\end{center}
%%%%%%%%%%%%%%%%%%%%%%%%%%%%%%%%%%%%%%%%%%%%%%%%%
%%%%%%%%%%%%%%%%%%%%%%%%%%%%%%%%%%%%%%%%%%%%%%%%%
In Figure \ref{FigureZLP}-a-b, we show the electron energy spectrum measured in our UTEM for different values of the number of electrons per pulse. 
The probe current, i.e the current inside the TEM column, is half the current emitted by the nanotip.
In the following, all experimental data are given as a function of the number of electrons emitted by the nanotip.
The spectrum measured on the same instrument operated in conventional DC emission mode is given in inset.
The latter has a FWHM of 0.45 eV obtained using 5 $\mu$A of emission current.
For comparison, our simulations predicted an energy width smaller by 0.25 eV.
The spectral width computed in laser-driven mode lies between 0.65 and 0.8 eV whereas the smallest energy spread measured is slightly lower than 1.1 eV (see Fig. 3-a, 3-d and 5-b).
This difference between experiment and simulation is similar in conventional DC and laser-driven emission mode and can be explained by the energy resolution of our spectrometer which is an old Gatan PEELS 666 located in a non-optimized environment.
As predicted in the previous section, the energy spectrum measured in normal and laser-driven emission mode has a different shape and an opposite asymmetry.
Care must be taken that the experimental spectra are presented as a function of the energy losses and therefore flipped with respect to the ones computed in the previous section.
In laser-driven mode, it is broader with a minimum energy width of 1 eV measured with the smallest current.
Our results clearly indicate that the energy width in laser-driven emission mode increases from a minimum value of 1 eV for pulses with in average less than one electron to 1.7 eV for electron pulses with 10 electrons per pulse.

It is important here to insist on the fact that pulses with a larger number of electrons are obtained by increasing the laser power incident on the nanotip.
Therefore, the increase of the measured spectral width can in principle be the combination of two effects.
First, we have seen that increasing the energy density injected in the metallic nanotip yields longer electron pulses and larger energy distributions due to a stronger perturbation of the electron gas.
Second, a larger number of electrons in each pulse means that space-charge effects will have a stronger effect (see Figure \ref{FigureSP}-a-b).
In Figure \ref{FigureZLP}-c, we show the minimum electron spot size (Full-Width at Half-Maximum) that could be obtained as a function of the number of electrons per pulse.
These results show that the spot radius increases from 1.4 nm for single electron pulses to 2 nm for pulses with 20 electrons.
This increase in spot radius can not be explained by the chromatic aberrations of the objective lens and suggests that electron-electron repulsion plays a major role on the spatio-temporal properties of the ultrashort electron pulses even for a small number of electrons.

%%%%%%%%%%%%%%%%%%%%%%%%%%%%%%%%%%%%%%%%%%%%%%%%%
\subsection{Characterization of the electron pulse properties based on the detection of electron energy gains}
%%%%%%%%%%%%%%%%%%%%%%%%%%%%%%%%%%%%%%%%%%%%%%%%%

In a pump-probe experiment, the temporal resolution is given by the cross-correlation between the \textit{pump} pulse and the \textit{probe} pulse.
Although the interaction between a free electron and a photon is not possible in vacuum, such an interaction becomes possible in the vicinity of a nanostructure as the optical near-field confined at the surface of the nano-object allows to satisfy energy-momentum conservation laws \cite{garcia_de_abajo_optical_2010}.
This electron-photon interaction in the optical near-field enables the absorption or emission of several photons by the swift electron and manifests itself as additional sidebands in the electron energy spectrum \cite{abajo_electron_2008,barwick_photon-induced_2009,garcia_de_abajo_multiphoton_2010,park_photon-induced_2010}.
These electron energy gains are a unique tool not only to characterize ultrashort electron pulses but also to measure quantitatively electric fields in photonic or plasmonic nanostructures as the probability of photon absorption/emission by the electron depends on the electric field along the electron trajectory \cite{barwick_photon-induced_2009,yurtsever_subparticle_2012,pomarico_mev_2017}.
We have performed such experiments to characterize the electron pulses in our UTEM.

%%%%%%%%%%%%%%%%%%%%%%%%%%%%%%%%%%%%%%%%%%%%%%%%%
%%%%%%%%%%%%%%%%%%%%%%%%%%%%%%%%%%%%%%%%%%%%%%%%%
\begin{center}
\begin{figure}[htp]
\centering
\includegraphics[width=14cm,angle =0.]{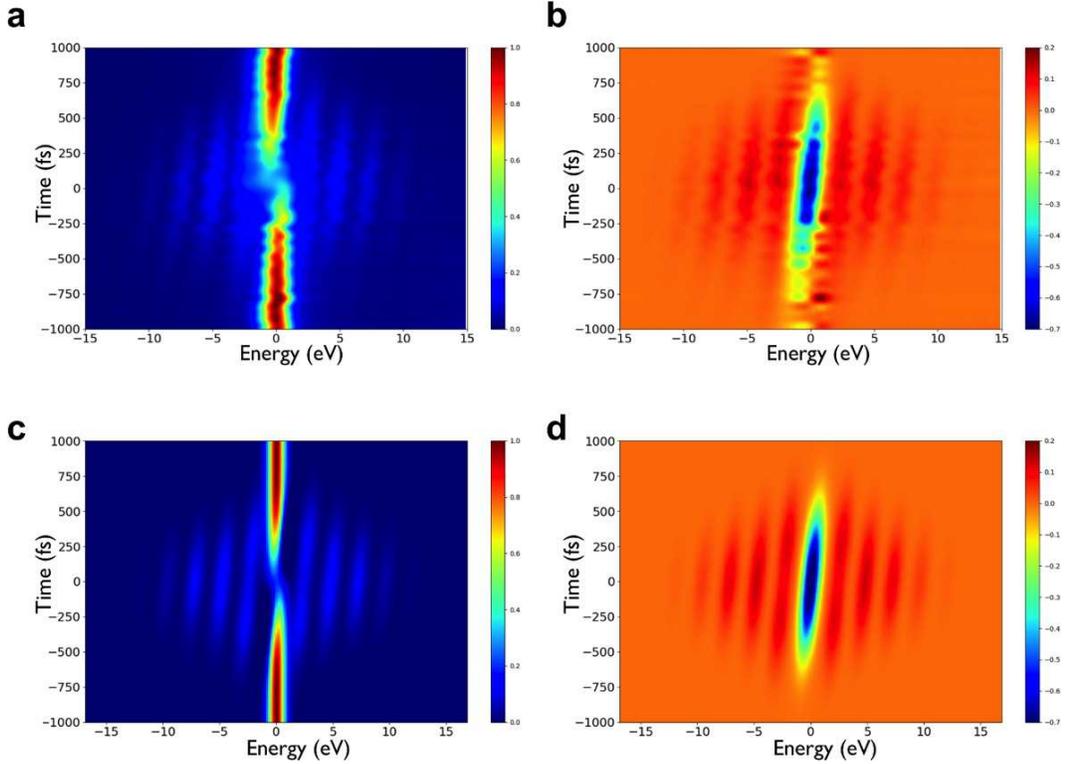}
\caption{(Color Online) a) Electron energy spectrum measured in the vicinity of a tungsten nanotip as a function of the delay between the electron pulse and the femtosecond laser pulse.
b) Difference between the spectra measured in a) and the electron spectrum measured without optical excitation. 
The data has been acquired with electron pulses having 8 electrons per pulse in the gun. The optical excitation wavelength was 515 nm (2.41 eV).
c) and d) Numerical simulations of the results presented in a) and b).}
\label{FigureEEGS}
\end{figure}
\end{center}
%%%%%%%%%%%%%%%%%%%%%%%%%%%%%%%%%%%%%%%%%%%%%%%%%
%%%%%%%%%%%%%%%%%%%%%%%%%%%%%%%%%%%%%%%%%%%%%%%%%

We have used a tungsten nanotip similar to the one used in the electron source of our UTEM to mediate the interaction between electrons and photons in the objective lens of the microscope.
The optical excitation is realized with the second harmonic of our laser at 515 nm.
The laser beam is focused on the nanotip using the light injection/collection system described above.
The delay between the laser and electron pulses is adjusted by a mechanical delay line.
Figure \ref{FigureEEGS}-a shows the electron energy spectrum as a function of the delay between the optical and electron pulses.
At long laser/electron delays, the electron energy spectrum is unchanged.
When the electron and laser pulses arrive simultaneously on the nanostructure, the electron/photon interaction strongly modifies the electron spectrum with several sidebands appearing on the gain and loss sides.
Using the theoretical formalism detailed in Park \textit{et al}, we have performed numerical simulations to analyse our experimental results and assess quantitatively the duration and energy width of the electron pulses \cite{park_photon-induced_2010}.
The details of these computations will not be given here for the sake of clarity but the interested reader can find all necessary information in the publication by Park et al.
We have modelled the tungsten nanotip as a nanocylinder with radius R=200 nm.
The optical near-field in the vicinity of the nanocylinder is computed from Mie theory and used to compute the absorption/emission probabilities as detailed in reference \cite{park_photon-induced_2010}.
Figure \ref{FigureEEGS}-c-d show examples of the simulated time-dependent electron energy spectra which are in good agreement with the experiments assuming an electron pulse duration of 400 fs and an incident laser intensity of 7.5 $10^8 \mathrm{W/cm}^2$ on the sample tip.
Furthermore, as can be clearly seen on Figure \ref{FigureEEGS}-a-b, the sidebands are tilted revealing a correlation between energy and time or chirp of the electron pulses  \cite{park_chirped_2012,Feist2015}.
This is related to the fact that the electrons within a pulse have different speeds due to the already discussed energy spread. 
By varying the delay between the photon and electron pulse, electrons of different speeds interact inelastically  with the sample. 
The absolute energy of the replicas are therefore different for different delays, but the relative energy differences are identical, leading to a
replica pattern tilted along the time axis.

%%%%%%%%%%%%%%%%%%%%%%%%%%%%%%%%%%%%%%%%%%%%%%%%%
%%%%%%%%%%%%%%%%%%%%%%%%%%%%%%%%%%%%%%%%%%%%%%%%%
\begin{center}
\begin{figure}[htp]1
\centering
\includegraphics[width=9cm,angle =0.]{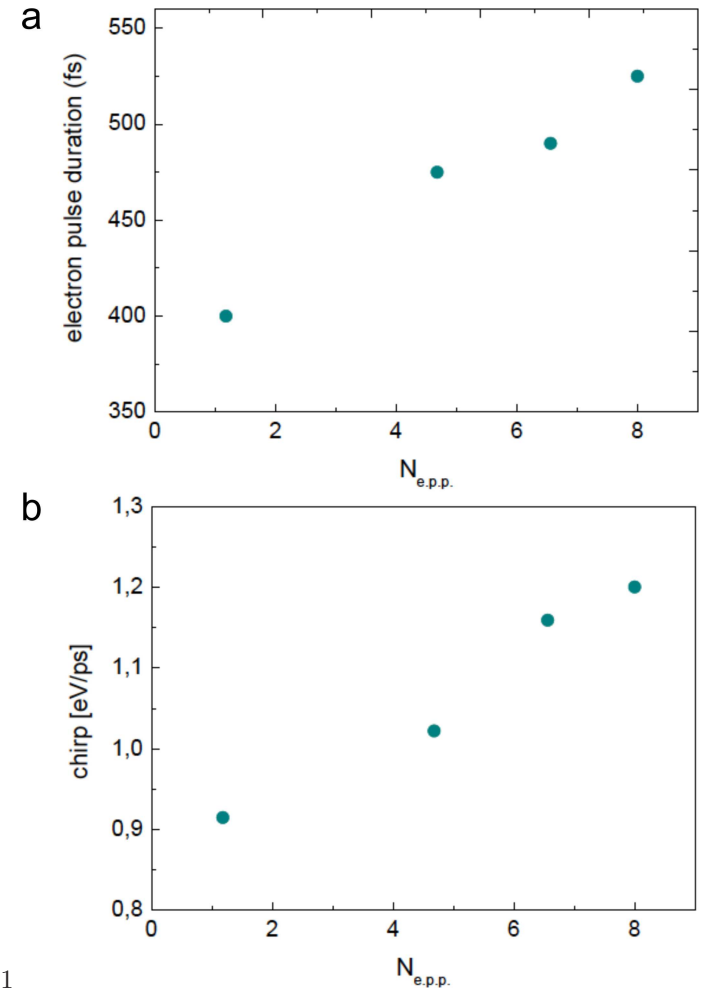}
\caption{(Color Online) a) Electron pulse duration determined from systematic EEGS experiments as a function of the number of electrons per pulse emitted in the electron gun.
b) Chirp of the electron pulses as a function of the number of electrons per pulse.}
\label{FigureBilanEEGS}
\end{figure}
\end{center}
%%%%%%%%%%%%%%%%%%%%%%%%%%%%%%%%%%%%%%%%%%%%%%%%%
%%%%%%%%%%%%%%%%%%%%%%%%%%%%%%%%%%%%%%%%%%%%%%%%%

Figure \ref{FigureBilanEEGS} shows the electron temporal FWHM and chirp as a function of the number of electrons per pulse.
These values yield the best agreement between our measurements and experiments for the complete spectro-temporal maps as well as for the energy-integrated temporal profiles and time-integrated spectra.
The minimum pulse duration currently attainable in our microscope is 400 fs.
It is obtained for electron pulses having in average less than one electron per pulse.
Our results show that an increase of the number of electrons per pulse is accompanied by an increase of the duration and chirp.
This observation together with the measured increase of the spot radius reveals that interparticle coulomb repulsion affects the properties of the femtosecond electron pulses already for very small number of electrons per pulse \cite{bach_coulomb_2019}.

%%%%%%%%%%%%%%%%%%%%%%%%%%%%%%%%%%%%%%%%%%%%%%%%%
%%%%%%%%%%%%%%%%%%%%%%%%%%%%%%%%%%%%%%%%%%%%%%%%%
\section{Applications of high-brightness Transmission Electron Microscopes}
%%%%%%%%%%%%%%%%%%%%%%%%%%%%%%%%%%%%%%%%%%%%%%%%%
%%%%%%%%%%%%%%%%%%%%%%%%%%%%%%%%%%%%%%%%%%%%%%%%%

In the previous section we have shown that our laser-driven cold-field emission source delivers ultrashort electron pulses with a duration of 400 fs and an energy spread close to 1 eV.
We have demonstrated that these ultrashort electron pulses can be focused in a nanometric spot.
The development of an ultrafast TEM with a high-brightness femtosecond electron source opens exciting avenues in nano-optics and nanomechanics for instance.
In the following, we discuss some of the applications of this instrument.

\subsection{High spatial resolution electron spectroscopy and cathodoluminescence}

Until now, most Photon Induced Near-field Electron Microscopy (PINEM) experiments performed to map the optical near-field of nano-objects have been performed with flat-photocathode-based UTEMs \cite{barwick_photon-induced_2009,piazza_simultaneous_2015,barwick_photonics_2015,pomarico_mev_2017}. 
The possibility offered by the recently developed high-brightness Schottky Field Emission Guns or Cold-field Emission Guns to focus femtosecond electron pulses in a nanometric spot has interesting consequences for the investigation of the optical responses of nanostructures.
For instance, this will allow to map the electric field by PINEM in confined geometries and investigate ultralocal modifications of the optical excitations supported by photonic or plasmonic nanostructures due to tiny modifications of their dielectric environment \cite{bosman_surface_2013,mazzucco_ultralocal_2012}.
The possibility to excite cathodoluminescence (CL) from nanoscopic systems will enable investigations of carrier dynamics in quantum-confined systems analog to time-resolved photoluminescence studies of semiconductor nanostructures \cite{meuret_lifetime_2016}.
The combination of these techniques (PINEM and CL) will be an powerful tool to explore the optical response of hybrid systems coupling a quantum emitter and a photonic nanostructure.

\subsection{Electron holography performed with pulsed electrons beam: towards quantitative field mapping at the femtosecond time scale}

\begin{center}
\begin{figure}[htp]
\centering
\includegraphics[width=12cm,angle =0.]{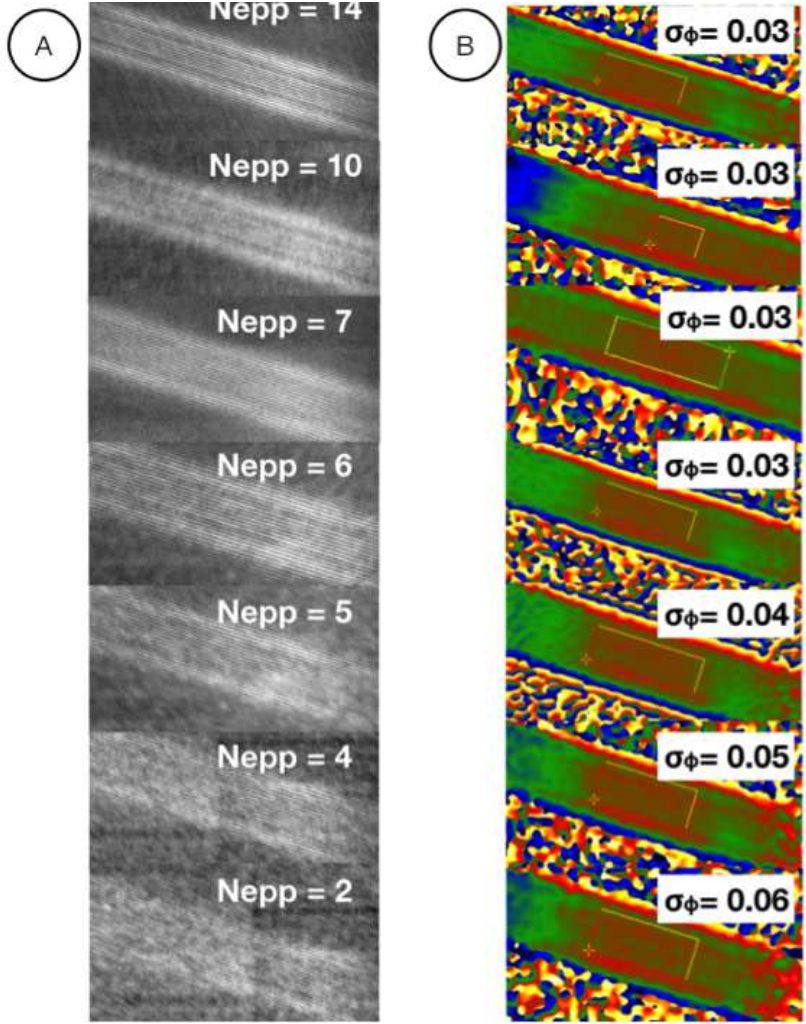}
\caption{(Color Online) a) Electron holograms recorded in the vacuum using ultrashort electron pulses as a function of the number of electrons per pulse $\mathrm{N}_{epp}$.
b)  Extracted electrons phase and measure of the phase standard deviation. Experimental conditions: 150 keV electrons, $\mathrm{f}_{laser}=2$ MHz, binning 2, Magnification =400 kX, Biprism voltage = 30 V, Stack of 90 holograms acquired with 20 s of individual exposure time \cite{houdellier_optimization_2019}.}
\label{FigureHolo}
\end{figure}
\end{center}

Proposed by Gabor as a new experimental method to improve the electron microscope resolution \cite{gabor_new_1948}, electron holography is a fully quantitative technique used to retrieve the electron wave phase recorded in an interference pattern \cite{dunin-borkowski_chapter_2015}. This pattern originates from the coherent overlap of the sample wave and a reference one. Digitally-recorded holograms can be used to extract the electron phase and locally map electric, magnetic, and strain fields \cite{hytch_nanoscale_2008}. The signal-to-noise ratio of phase measurement will directly influence the precision of the measured quantity. 
In a recent publication, we reported on the necessary optimization of electron holography experiments acquired with femtosecond electron pulses due to the low electron probe current available in our laser driven nanoemitter technology \cite{houdellier_development_2018,houdellier_optimization_2019}. We showed that working using such an ultrafast electron source is ultimately equivalent to work in severe low-dose imaging conditions with a dose lying between 5-20 electrons per square Angstr\"{o}m. This configuration leads to difficulties especially from the need to increase the exposure time required for the acquisition of exploitable electron holograms.
We have then optimized the experimental parameters to maximize the contrast of the hologram fringes acquired with femtosecond electron pulses showing that exposure times of 100 s or more are necessary to record fringes with enough contrast.
Indeed, in theory, increasing the number of electrons contributing to the hologram improves the signal to noise ratio of the reconstructed phase. However, in reality, long exposure times are prohibited by instabilities of the experimental set-up during acquisition. Using continuous electron beam it is well known that exposure times are limited to a few seconds to record a good hologram, reaching few tens of seconds in a particularly stable environment and conditions. The two major sources of instabilities are the drift in the position of the hologram fringes and the specimen position. As a consequence, using more than 100 s of exposure time prevents us to record hologram with sufficient contrast and field of view.  
The solution to increase the exposure time without deteriorating the hologram contrast, currently implemented using continuous electron beam, is to record series of holograms in \textit{image stacks}. The holograms are realigned in advanced post-processing before integrating the information.  In this way exposure times equivalent to about 400 s \cite{voelkl_approaching_2010} and 900s \cite{mcleod_phase_2014} have been obtained in medium-resolution electron holography, and 60 s for high-resolution electron holography where drift is more critical \cite{niermann_averaging_2014}.
We have implemented this solution using electron holograms acquired with ultrashort electrons pulses achieving 30 min of acquisition time \cite{houdellier_optimization_2019}. This solution allowed us to record optimized electrons hologram yielding small phase standard deviation and a phase detection limit in the 100 mrad range with 3 nm spatial resolution. 
Recently using electron holography experiments, we have shown that the spatial coherence of the ultrashort electron beam is not affected by the numbers of electrons per pulse. Using hologram stacks techniques described in \cite{houdellier_optimization_2019}, we have recorded 90 electrons holograms of 20 s of exposure time, and determined the standard deviation of the reconstructed phase for each condition. Figure \ref{FigureHolo} reports this result confirming that, regarding electron holography experiment, a number of electrons per pulse between 10 and 20 seems to be the optimum. Above this value, tip damage can occur. 

%%%%%%%%%%%%%%%%%%%%%%%%%%%%%%%%%%%%%%%%%%%%%%%%%
%%%%%%%%%%%%%%%%%%%%%%%%%%%%%%%%%%%%%%%%%%%%%%%%%
%%%%%%%%%%%%%%%%%%%%%%%%%%%%%%%%%%%%%%%%%%%%%%%%%
%%%%%%%%%%%%%%%%%%%%%%%%%%%%%%%%%%%%%%%%%%%%%%%%%

The fact that the standard deviation on the phase of the electron hologram is independent of the number of electrons per pulse shows that the spatial coherence of the beam in the electron hologram field of view is not affected by the coulomb repulsion. This can be easily understood thanks to the specific illumination condition conditions used for the electron holography experiments. Indeed, as in conventional electron holography, the illumination is strongly demagnified perpendicular to the biprism wire \cite{volkl_introduction_1999}. Therefore, the electron spot broadening observed experimentally and arising from the electron-electron repulsion inside the pulse, will not play any role in the hologram area. On the other hand, the consequence of this demagnification is a strong decrease of the electrons dose \cite{houdellier_development_2018,houdellier_optimization_2019}.
Now that we have properly determined the optimum experimental conditions, the next step will be to perform ultrafast electron holography using the pump-probe setup to study the dynamics of electric and magnetic fields in nanoobjects \cite{arbouet_chapter_2018}. Another interesting perspective is to perform dark field electron holography experiments in order to map the strain fields associated to the mechanical vibrations of nanostructures \cite{nelet_acoustic_2004}.
The main advantage of holography in all these studies arises from the intrinsic quantitative nature of the results obtained. 

\subsection{Analysis of the structural dynamics of nanomaterials using ultrafast electron diffraction}

The capabilities to combine real space and reciprocal space information is one of the major advantages of the Transmission Electron Microscope. By simply changing the excitation of intermediate and projectors lenses the instrument can record electron diffraction micrograph on a very localized area of a nanocrystal. Furthermore, the strength of electron diffraction performed in a TEM lies in the various modes available depending on the illumination conditions selected through the excitation of the condensor lenses and the selection of apertures. By using a parallel beam combined with an appropriate selected-area aperture, point diffraction patterns can be recorded from which the crystal symmetry and quantitative information such as lattice parameters can be retrieved \cite{williams_transmission_1996}. These patterns called Selected Area Electron Diffraction (SAED) patterns are also fundamental for the implementation of all conventional analysis performed in a TEM such as bright field, dark field or weak beam imaging techniques. 
By using a convergent illumination beam, disk patterns named Convergent Beam Electron Diffraction (CBED) patterns can be recorded \cite{morniroli_large-angle_2002}. The diffracted intensities recorded inside these disks exhibit specific excess lines of electrons while the central transmitted disk contained the superimposition of all the associated deficiency lines. The profile of each line, usually called Rocking curve, can be used to quantitatively determine a wide range of information such as crystal thickness, structure factor, Debye Waller factor, while their positions can be used to determine the crystal strain state with a very high precision [14]. By increasing the incident convergence angle, Kossel-M\"{o}llensted patterns are obtained. They are equivalent to CBED patterns but somehow less useful due to the superimposition of (hkl) excess line with (-h-k-l) deficiency lines coming from these large angle conditions. In order to take advantage of such high convergence beam, excess lines can be filtered out by the selected aperture using specific defocused conditions. These patterns known as LACBED (Large Angle CBED) exhibit the superimposition of deficiency lines over a wide angle together with the shadow image of the sample due to the defocused conditions \cite{morniroli_cbed_2006}. LACBED can be used to study crystal symmetry, as well as local crystal defects such as dislocations or stacking faults.
All these well-known configurations can be adapted using our coherent UTEM, but, as previously described, due to the pulsed illumination the emitted current is strongly reduced \cite{houdellier_development_2018}. Indeed, we have seen previously that an emission current in the picoampere range is typically extracted in our ultrafast CFEG using 2 MHz of laser repetition rate. 
This very low current must be compared to the ones in the microampere range usually used in continuous CFEG emission. These 6 orders of magnitude in current between a standard FE-TEM and our FE-UTEM has a direct consequence on the electron diffraction capabilities, as in electron holography and all other techniques implemented. Indeed, all UTEM techniques can be described as low dose techniques and will then need high exposure times to generate patterns with enough signal over noise. The sources of instabilities during the exposure time, as well as the properties of the recording camera such as Detective Quantum Efficiency (DQE) and Point Spread Function (PSF), will have a crucial role on the results quality. 

\begin{center}
\begin{figure}[htp]
\centering
\includegraphics[width=12cm,angle =0.]{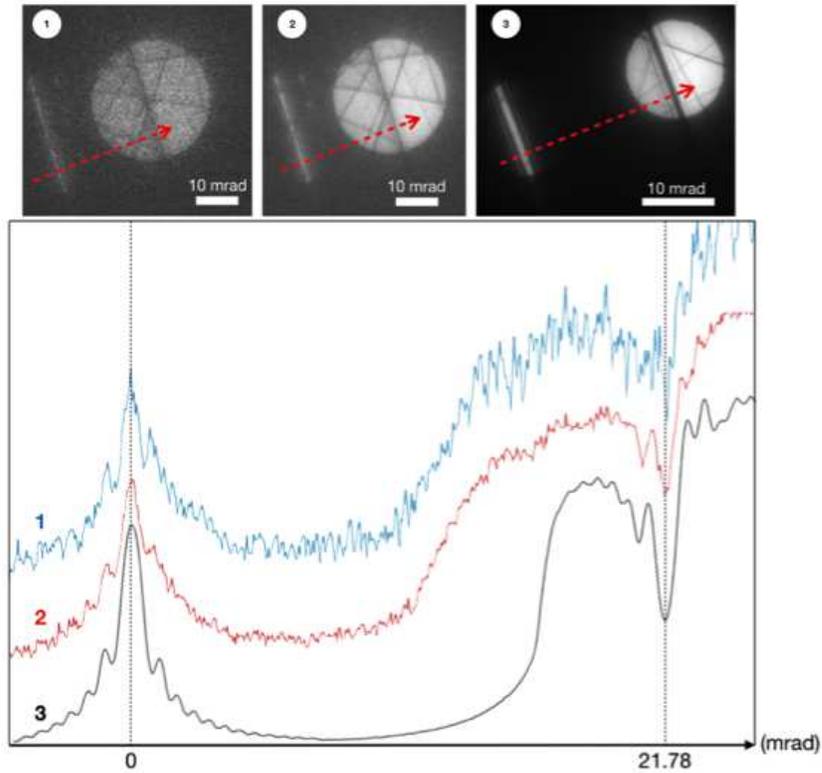}
\caption{(Color Online) Two beam (004) CBED patterns and rocking curve profiles acquired on a Silicon lamella with 1) 20 s of exposure time using ultrashort electron pulses, 2) 30 min of exposure time, thanks to a stack of 90 CBED patterns of 20 s of individual exposure time, using ultrashort electron pulses and 3) 0,5 s of exposure time using a continuous electron beam. Experimental conditions of the ultrafast electron source : 150 keV electrons, $\mathrm{P}_{laser} = 8$ mW, $\mathrm{N}_{epp}=6$,  $\mathrm{f}_{laser}=2 $MHz.}
\label{FigureDiff1}
\end{figure}
\end{center}

Figure \ref{FigureDiff1} reports examples of (004) two-beams CBED patterns obtained on a Silicon lamella. The rocking curve has been extracted and compared to the one obtained using a continuous electron beam on the same area. While the (004) rocking curve acquired with continuous beam can be extracted with a good contrast after 0.5 s of exposure time only, the signal to noise ratio (S/N) of the one extracted after 20 s in pulsed conditions is not sufficient to be practically useful. In order to improve the patterns acquired using ultrashort electrons pulses, stack acquisition of CBED patterns will need to be implemented as already done for electron holograms \cite{voelkl_approaching_2010}. After a total of 30 min of exposure time obtained using a stack of 90 CBED patterns of 20 s of individual exposure time, an interpretable rocking curve is obtained but the S/N remains much worse than the one obtained with the continuous emission condition. In figure \ref{FigureDiff1} we can then notice an improvement of the rocking curve S/N thanks to the stack, even if the fringes contrast still remains not comparable to the one obtained using continuous emission. Nevertheless, the signal is sufficient to extract quantitative information such as the thickness. A value of 270 nm has been measured using this pattern. 

%%%%%%%%%%%%%%%%%%%%%%%%%%%%%%%%%%%%%%%%%%%%%%%%%
%%%%%%%%%%%%%%%%%%%%%%%%%%%%%%%%%%%%%%%%%%%%%%%%%
%%%%%%%%%%%%%%%%%%%%%%%%%%%%%%%%%%%%%%%%%%%%%%%%%
%%%%%%%%%%%%%%%%%%%%%%%%%%%%%%%%%%%%%%%%%%%%%%%%%

\begin{center}
\begin{figure}[htp]
\centering
\includegraphics[width=12cm,angle =0.]{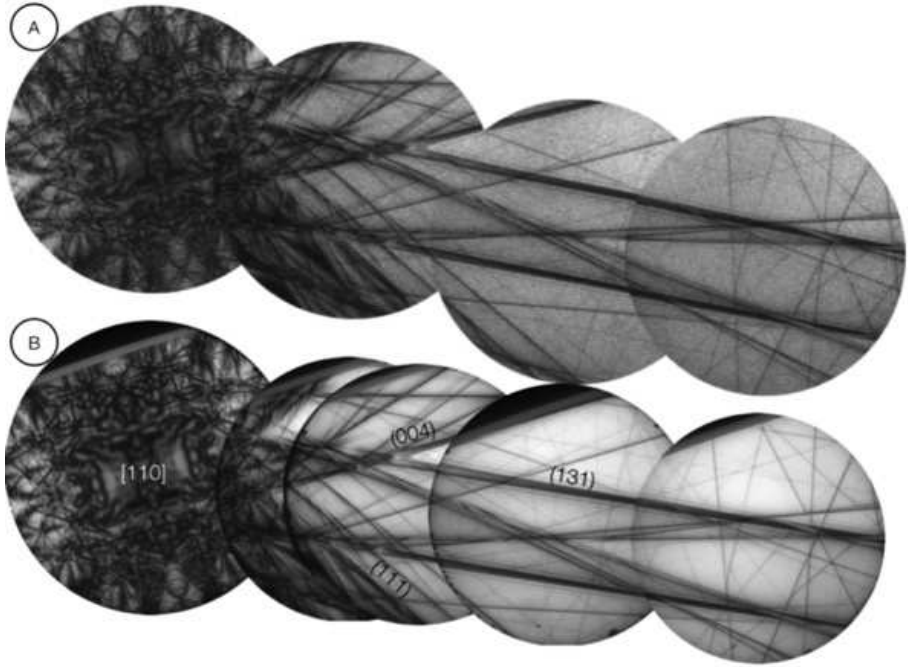}
\caption{(Color Online) LACBED patterns obtained on a Si lamella under the same conditions using ultrashort electrons pulses and 150 s of exposure time (a) and continuous electron beam and 0,5 s of exposure time (b). Experimental conditions of the ultrafast electron source : 150 keV electrons, $\mathrm{P}_{laser} = 8$ mW, $\mathrm{N}_{epp}=6$,  $\mathrm{f}_{laser}=2$MHz.}
\label{FigureDiff2}
\end{figure}
\end{center}

Filtering the recorded pulsed CBED pattern from all the inelastic signal using an imaging filter will certainly improve the quality of the extracted rocking curve signal, as it is usually performed in conventional quantitative CBED analysis \cite{zuo_direct_1999}.  Such a device is not installed in our experimental setup \cite{houdellier_development_2018}, but LACBED patterns is an alternative solution. Indeed, using small SA aperture and a large defocus enables to filter out the inelastically scattered electrons in momentum space instead of energy space. 
Figure \ref{FigureDiff2} reports two series of LACBED patterns recorded in the [110] zone axis orientation. Figure \ref{FigureDiff2}-a has been obtained using pulsed emission and 150 s of exposure time. On the same area Figure \ref{FigureDiff2}-b has been recorded using continuous emission and 0.5 s of exposure time. 
We can notice that high quality LACBED patterns can be obtained despite the very low emission current provided by the ultrafast electron source. Thanks to the high quality of these patterns, the crystal symmetry and deficiency lines positions can be easily extracted. 

%%%%%%%%%%%%%%%%%%%%%%%%%%%%%%%%%%%%%%%%%%%%%%%%%
%%%%%%%%%%%%%%%%%%%%%%%%%%%%%%%%%%%%%%%%%%%%%%%%%
%%%%%%%%%%%%%%%%%%%%%%%%%%%%%%%%%%%%%%%%%%%%%%%%%
%%%%%%%%%%%%%%%%%%%%%%%%%%%%%%%%%%%%%%%%%%%%%%%%%

Now that we have properly determined the optimum conditions to acquire electron diffraction patterns using ultrashort electron pulses, especially in convergent beam illumination, our goal is to study the dynamics of nanostructures using our ultrafast TEM. These experiments are still under progress and will be the subject of dedicated publications. 
However, we can already state that, by fitting the ultrafast rocking curves obtained using stacks of CBED patterns and performing numerical simulations based on the dynamical theory of high-energy electrons for different pump-probe delays, we should be able to determine the evolution of the crystal complex structure factor. Ultimately this information can be used to refine atomic positions and charge density in complex materials as a function of time. Finally, thanks to the high quality LACBED patterns and due to their three-dimensional nature, we could extract the full time-dependent local deformation gradient tensor by studying the dynamical evolution of different HOLZ deficiency lines positions. This should allow us to image phonon modes up to the terahertz regime and become an ideal tool for the further development of nanophononics for instance.
First studies of the dynamics of acoustic waves by ultrafast convergent-beam diffraction using a flat-photocathode have been performed in the cases of a silicon wedge \cite{yurtsever_kikuchi_2011} and graphite nanoslab \cite{liang_observing_2014}.
Results in the same direction have recently been reported by the G\"{o}ttingen group using Kossel-M\"{o}llensted patterns in graphite samples thanks to their Schottky based UTEM \cite{feist_nanoscale_2018}.

%%%%%%%%%%%%%%%%%%%%%%%%%%%%%%%%%%%%%%%%%%%%%%%%%
%%%%%%%%%%%%%%%%%%%%%%%%%%%%%%%%%%%%%%%%%%%%%%%%%
\section{Conclusion}
%%%%%%%%%%%%%%%%%%%%%%%%%%%%%%%%%%%%%%%%%%%%%%%%%
%%%%%%%%%%%%%%%%%%%%%%%%%%%%%%%%%%%%%%%%%%%%%%%%%
We have developed an ultrafast TEM based on a laser-driven cold-field emission source.
This electron source delivers ultrashort electron pulses with a duration of 400 fs and an energy spread close to 1 eV.
The spectral and temporal properties of the electron pulses result from the combined influence of the electron ultrafast dynamics after the excitation by the laser pulses and propagation effects which alter their propagation from the source to the sample and detectors.
The high brightness of this electron source opens immense possibilities for the study of the optical, mechanical and magnetic properties of nano-objects.

%%%%%%%%%%%%%%%%%%%%%%%%%%%%%%%%%%%%%%%%%%%%%%%%%
%%%%%%%%%%%%%%%%%%%%%%%%%%%%%%%%%%%%%%%%%%%%%%%%%
\section{Acknowledgements}
%%%%%%%%%%%%%%%%%%%%%%%%%%%%%%%%%%%%%%%%%%%%%%%%%
%%%%%%%%%%%%%%%%%%%%%%%%%%%%%%%%%%%%%%%%%%%%%%%%%

The authors thank the \textit{Institut de Physique du CNRS} and \textit{Agence Nationale de la Recherche} for financial support (ANR grant ANR-14-CE26-0013).
This work was supported by \textit{Programme Investissements d'Avenir} under the program ANR-11-IDEX-0002-02, reference ANR-10-LABX-0037-NEXT (\textit{MUSE} grant).
This work was supported by the computing facility center CALMIP of the University Paul Sabatier of Toulouse.
The authors acknowledge financial support from the European Union under the Seventh Framework Program under a contract for an Integrated Infrastructure Initiative
(Reference 312483-ESTEEM2). The authors are grateful to J. D. Blazit and M. Pelloux for their contribution to the light injector design and fabrication, R. Cours for sample preparation and E. Snoeck and M. J. H\"{y}tch for their support.

\section{References}

\bibliographystyle{unsrt}
\bibliography{biblio.bib}

\end{document}